\documentclass[referee]{aa}
\usepackage{graphicx,epsf}
\newcommand{\AJ}{AJ}
\newcommand{\ARAA}{ARA\&A}

\newcommand{\AaA}{A\&A}

\newcommand{\ApJ}{ApJ}
\newcommand{\MNRAS}{MNRAS}
\newcommand{\Natur}{Nature}

\newcommand{\PhRvD}{Phys. Rev. D}
\newcommand{\PhRvL}{Phys. Rev. Lett.}
\newcommand{\PhR}{Phys. Rep.}

\begin{document}
\title{Cosmic Microwave Background anisotropies beyond the third peak}
\author{N. Aghanim\inst{1}\inst{,3} \and P. G. Castro\inst{2} 
\and A. Melchiorri\inst{2} \and J. Silk\inst{2}}
\institute{IAS-CNRS, Universit\'e Paris Sud, B\^atiment 121, F-91405 Orsay 
Cedex \and Denys Wilkinson Building, University of Oxford, 
Keble Road, Oxford, OX 3RH, UK \and IAP-CNRS, 98 bis Boulevard Arago, F-75014 
Paris} 
\date{Received date / accepted date}
\abstract{Forthcoming CMB experiments will allow us to accurately investigate 
the power spectrum at very small scales ($\ell > 1000$). 
We predict the level of the primary anisotropies,
given the actual CMB measurements. The secondary anisotropies 
generated after  matter-radiation decoupling contribute  
additional power in the tail of the CMB power spectrum. 
Together with the primary anisotropies, we compute the 
predicted power spectra for three dominant secondary effects induced
by photon scattering. We predict these secondary contributions in flat 
cosmological models for parameters 
in agreement (to $2\sigma$) with the values allowed by current 
parameter estimates.
}
\maketitle
\markboth{Cosmic Microwave Background anisotropies beyond the third peak}{}
\section{Introduction}

The last several  years have been an exciting period for observational
cosmology, particularly in the field of the Cosmic Microwave Background
(CMB) research. With  recent CMB balloon-borne 
and ground-based experiments such as TOCO \cite[]{miller99}, 
BOOMERanG \cite[]{de_bernardis2000,mauskopf2000,netterfield2001},
MAXIMA \cite[]{hanany2000,lee2001,stompor2001}, and DASI 
\cite[]{halverson2001}, we are entering a new era of 
precision cosmology that enables us to use the CMB anisotropy measurements 
to constrain the cosmological parameters and the underlying theoretical 
models. From all of these experiments, a firm detection of
the first peak in the CMB anisotropy angular power spectrum has now been
obtained. 
Moreover, in the framework of adiabatic cold dark matter (CDM) models, the
position, amplitude and width of the measured peak provide strong evidence 
for the inflationary predictions of a low curvature (flat) universe and 
a scale-invariant primordial spectrum \cite[]{dodelson2000,melchiorri2000}.
Furthermore, the latest results from BOOMERanG \cite[]{netterfield2001}
and DASI \cite[]{halverson2001} point to the presence of 
second and third peaks, confirming the theoretical
prediction of acoustic oscillations in the primeval plasma 
and shedding new light on various cosmological and 
inflationary parameters \cite[]{de_bernardis2002,pryke2001,wang2001}.
The locations and amplitudes of three acoustic peaks and two dips in the 
last releases of the CMB data have been determined in a model-independent way
in recent work (see e.g. \cite{durrer2001}). 
In general, the location and amplitude of the 
first acoustic peak is determined at more than $3 \sigma$ confidence. 
The next two peaks and dips are determined at a confidence 
level above $1 \sigma$ but below $2 \sigma$. 
It is however worth mentioning the work of \cite{miller2001}, 
which finds that the 
second peak is a $2 \sigma$ detection and that the third peak is not 
detected to any reasonable significance.

The next generation of CMB experiments will be even more powerful 
as they are designed to achieve, down to  arcminute scales, 
accurate measurements of both the
temperature anisotropies and polarised emission of the CMB. Through 
these data,
we expect multiple oscillations in the spectrum to 
be unambiguously detected and, thanks to the polarisation measurements, 
this should 
enable us to remove some of the degeneracies that are still affecting 
parameter estimation.

However, the present CMB data, as we will see in one of the 
following sections, already constrain the shape of the 
angular power spectrum of 
primary anisotropies with good accuracy. 
The present poor determination of some of
the cosmological and astrophysical parameters is more
 related to intrinsic degeneracies 
(different models leading to the same spectrum) rather than to 
the precision of the actual CMB measurements. 
Given this uncertainty, we can forecast in a reliable way
the level of {\it primary} anisotropies on arcminute and 
sub-arcminute scales, where the effect of {\it secondary} 
anisotropies (i.e. produced well after matter-radiation decoupling) 
should start to dominate and hence significantly affect the CMB 
measurements.
The secondary anisotropies can be generated due to photon interactions 
with the matter
potential wells for example in the Rees-Sciama effect \cite[]{rees86},
lensing \cite[]{seljak96b} and the  ``butterfly'' effect
\cite[]{birkinshaw83}. The other secondary
anisotropies are induced by the interaction of CMB photons with free
electrons such as in the Sunyaev-Zel'dovich (SZ) effect
\cite[]{suniaev80}, the Ostriker-Vishniac (OV) effect
\cite[]{ostriker86,vishniac87}, and form the effects of 
early inhomogeneous
reionisation \cite[]{aghanim96,gruzinov98,knox98}. 

In the present work, we first forecast the level of primary anisotropies
on arcminute and sub-arcminute scales by comparing the 
recent CMB data with a template
of cosmological models (Section 2).
We then quantify, in Section 3, the contribution of the secondary scattering
effects that are likely the dominant contributions at small scales.
The contributions are computed for a set of cosmological parameters 
allowed by the present data (within $2\,\sigma$ error
bars). The corresponding power spectra are presented together
with our results in Section 4. Our discussion and conclusions are 
given in Section 5.
\section{Forecasting primary anisotropies}

Our goal in this section is to forecast the level of 
primary anisotropies on very small scales ($\ell > 1000$), 
given the actual CMB measurements that span an interval in
angular scales of $2 < \ell < 1000$.
The first step in our procedure consists of building a
template of cosmological models and in comparing each model 
with the CMB observations through a likelihood.
The theoretical models of our template are computed using the 
publicly available {\sc cmbfast} program. They are defined by
6 parameters sampled as follows: 
$\omega_{cdm}=\Omega_{cdm}h^2= 0.01,...0.40$ (step $0.01$); 
$\omega_{b}=\Omega_bh^2 = 0.001, ...,0.040$ (step $0.001$); 
$\Omega_{\Lambda}=0.0, ..., 1.0$ (step $0.05$) and 
$\Omega_k$ such that $\Omega_m=0.1, ..., 1.0$ (step $0.05$).
The value of the Hubble constant is not an independent
parameter, since 
$h=\sqrt{(\omega_{cdm}+\omega_b)/(1-\Omega_k-\Omega_\Lambda)}$.
We vary the spectral index of the primordial density perturbations
within the range $n_s=0.60, ..., 1.40$ (step $0.02$) 
and we rescale the amplitude of fluctuations by a
pre-factor $C_{10}$, in units of $C_{10}^{COBE}$, with 
$0.50 < C_{10} < 1.40$.

The models of the template are then compared with the recent 
BOOMERanG-98, DASI and MAXIMA-1 results.
The power spectra from these experiments were estimated in 
$19$, $9$ and $13$ bins respectively, spanning the range
$25 \le \ell \le 1150$. 
For the DASI and MAXIMA-I experiments, we use the publicly available
correlation matrices and window functions.
For the BOOMERanG experiment, we assign a flat space 
for the spectrum in each bin $\ell(\ell+1)C_{\ell}/2\pi=C_B$; 
we approximate the signal $C_B$ inside
the bin to a Gaussian variable and we consider $\sim 10 \%$ 
correlations between the various bins.
We consider $10 \%$, $4 \%$  and $5 \%$ Gaussian distributed 
calibration errors (in $\mu K$) for the BOOMERanG-98, DASI 
and MAXIMA-1 experiments respectively and 
we included the beam uncertainties by the analytical marginalisation
method presented in \cite{bridle2001}.
We also include the COBE data using Lloyd Knox's RADPack packages.
The likelihood for a given cosmological model is then defined by 
$-2{\rm ln} {\cal L}=(C_B^{th}-C_B^{ex})M_{BB'}(C_{B'}^{th}-C_{B'}^{ex})$
where  $M_{BB'}$ is the Gaussian curvature of the likelihood 
matrix at the peak. 
We then plot in the $\ell(\ell+1)C_{\ell}-\ell$ plane
 the envelope of all the models that are consistent at 
$95 \%$ c.l. with the CMB data (Fig. \ref{fig:contrib}). This 
provides us with
the predicted power spectrum of the primary anisotropies as deduced from
the actual data. We consequently define above this band
the region where, at $95 \%$ c.l., a contribution from primary 
anisotropies is {\it unlikely} to be present.
In performing the envelope of all the models,
we have neglected the correlations between different $\ell$ modes
for a single model: This simple procedure is also
the most conservative approach. 
Moreover, our computations rely on two main assumptions:
\begin{itemize}
\item The primary CMB can be described by one of 
the theoretical models in our template (i.e.
we do not consider possible deviations such as  scale-dependent
spectral indexes, isocurvature primordial perturbations or topological 
defects).
\item The present CMB data are not affected by unknown 
foregrounds or systematics.
\end{itemize}
The first assumption is a theoretical prior mainly supported 
by the good agreement between the theoretical template and the present 
CMB data. The same underlying model seems also to fit  the shape of 
the matter power spectrum 
on small scales obtained, for example, from recent galaxy surveys. 
The second assumption is based on the numerous tests performed by the
experimental teams to estimate the Galactic contamination in the so-called
CMB channels (150 or 30 GHz). It also relies, in particular,
on the point source extraction
carried out by the DASI team. Moreover, the overall 
consistency of the CMB data from these experiments that have sampled 
various regions of the sky, at different frequencies and with different
techniques (bolometres and radiometres) reinforces our confidence in this
second assumption. In the following, we will generalise the {\it standard} 
concept 
of foregrounds to include the secondary anisotropies that cannot be
spectrally disentangled from the primary anisotropies,
and thus corrected for, by the present 
experiments, and we will investigate their possible contributions.

\section{Secondary contributions}
An increasingly large number of studies has been devoted to the CMB
anisotropies in general, and the secondary anisotropies in particular. We 
refer the 
reader to \cite{hu2001} and \cite{white2002}. The latter give a rather 
comprehensive guide to the literature on the topics related to the CMB, 
including secondary anisotropies. 
In the present study, we concentrate on the dominant contributions 
due to secondary effects generated through
interactions with electrons. The induced temperature anisotropies 
contribute to the total signal at different angular scales. Between a 
few arcminutes to a few tens of arcminutes, the SZ effect dominates, whereas 
at smaller angular scales typically down to arcminute scales the 
extra power essentially comes from the OV effect. 
\subsection{Sunyaev-Zel'dovich effect}
The SZ effect (see \cite{birkinshaw99} for a recent review) is a 
well-studied effect which designates the inverse Compton interaction of the
CMB photons with the free electrons of the hot intra-cluster
medium. This effect is usually ``decomposed'' into two parts: The
so-called thermal SZ (TSZ), and the kinetic SZ (KSZ) effects.  The TSZ
amplitude is a measure of the pressure integral along the line of
sight. It is directly linked to the intra-cluster gas properties
(temperature $T_e$ and optical depth $\tau$). It has a very peculiar
spectral signature that translates into an excess of brightness at
high frequencies (850 $\mu$m) and a decrement at low frequencies (2
mm). For hot clusters, it can be modified to include relativistic
corrections (see \cite{rephaeli95} for example). The KSZ effect
corresponds to a first-order Doppler effect when a galaxy cluster
moves with respect to the CMB rest frame with a radial peculiar
velocity $v_r$. The KSZ anisotropies have the same spectral signature
as the primary CMB anisotropies.  \par
The SZ effect is one of the most important sources of
secondary anisotropies. Due to its spectral signature, the TSZ can in
principle be separated from the other components (in particular from
the CMB primary anisotropies) by means of accurate multi-wavelength
observations in the millimetre/submillimetre range
\cite[]{hobson98,bouchet99,snoussi2001}.  The only remaining
contribution is therefore that associated with the KSZ anisotropies
which cannot be separated from the primary fluctuations via
multi-wavelength measurements.  However, most of the present
and future CMB experiments due to their lack of frequency coverage
are unable to detect and subtract the TSZ anisotropies from the
measurements.  This contribution is thus likely to be present in the
measured signal. 
In the present study, we investigate the SZ anisotropies using a 
semi-analytical model, the most recent  version of which 
(assumptions and procedures) is described in detail in
\cite{aghanim2001}. It is based on the computation of predicted
cluster number counts from a Press-Schechter \cite[]{press74} mass 
function modified, according to \cite{sheth99} and \cite{wu2000}, to
better fit the numerical predictions of the number of dark matter halos 
in the mass
range $5.\,10^{13}$ to $10^{16}$ solar masses. Each cluster is physically
described by its intra-cluster gas temperature (assumed isothermal)
and electron density distribution (following the well-known 
$\beta$-profile, $\beta=2/3$). Provided with these quantities and the 
cluster peculiar velocities, both TSZ and KSZ anisotropies are computed 
for the cluster population. The SZ contribution (power spectrum) is then 
computed via simulated maps of both KSZ and TSZ components. \par
\begin{figure}[htb]
\begin{center}
\epsfxsize=12cm
\epsfysize=10cm
\epsffile{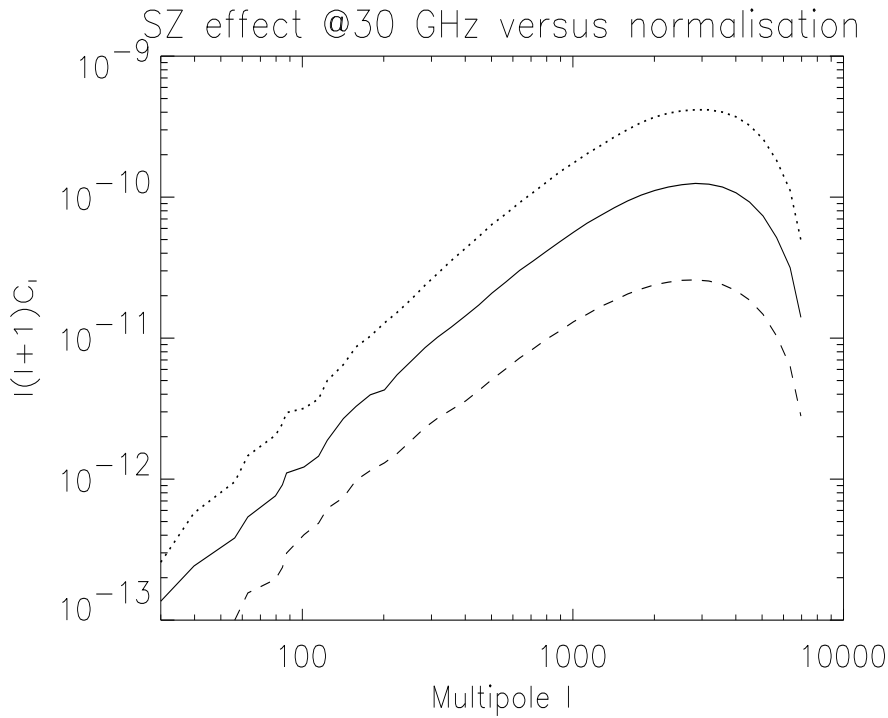}
\end{center}
\caption{The SZ power spectrum computed for the ``concordance model''
($h=0.71$, $\Omega_m=0.3$, $\Omega_\Lambda=0.7$). The three curves
represent the variation of the power spectrum with the normalisation
factor $\sigma_8=\sigma_8^0\Omega_m^{-0.47}$ (from \cite{viana99}).
The dashed line is obtained for $\sigma_8^0\,=\,0.4$, the solid line
for 0.5 and the dotted line for 0.6. }
\label{fig:sigm}
\end{figure}

Several techniques, based on analytic or 
semi-analytic computations and on numerical simulations, can be used 
to predict the SZ signal of a cluster population (as for example in
\cite{cooray2000}, \cite{da_silva2000}, \cite{komatsu2000},
\cite{refregier2000}, \cite{majumdar2001}, \cite{springel2001} 
 (and references therein). 
These authors give different predictions, especially for the power
spectrum. An illuminating comparison illustrating this point has
been performed by \cite{springel2001}. The most noticeable result of the
comparison is a scatter
of about one order of magnitude in the SZ power spectra at all angular
scales. This scatter is partially due to the differences between 
the different methods used for the computation, and to their intrinsic
limitations. We note, for example,
that the semi-analytic predictions apparently give somehow higher amplitudes 
than numerical simulations. As for the latter, they are limited by 
numerical resolution, which especially affects the results at small angular 
scales. In addition, the amplitude of the SZ contribution crucially 
depends on the physical assumptions entering in the computation. The
observed variations in the predicted power spectra can be due on the
one hand to the predicted number of structures and on the other hand to 
the physical description of the intra-cluster
gas, and its evolution with redshift. In this context, one of the most 
influencial parameters for the number count computations is the 
normalisation factor $\sigma_8$ which is also rather uncertain. The 
amplitude of the SZ power spectrum 
can indeed vary by an order of magnitude with rather small variations of 
$\sigma_8$. This dependence is illustrated in figure 
\ref{fig:sigm} where we use the expression given in \cite[]{viana99}: 
$\sigma_8=\sigma_8^0\Omega_m^{-0.47}$ with $\sigma_8^0\,=\,0.4$ (dashed 
line), 0.5 (solid line) and 0.6 (dotted line), and $\Omega_m=0.3$. 
It is therefore important to realise  that due to our present limited
understanding of the structure formation and evolution (especially for 
the gaseous part), the predicted SZ power spectra only provide an order 
of magnitude estimate of the effect rather than exact and precise amplitudes.
\medskip

\subsection{Inhomogeneous reionisation}
It has become common knowledge that the universe underwent 
reionisation that was completed around a redshift $z_i\sim 6$. The 
evidence for reionisation relies on observations of the Gunn-Peterson
\cite[]{gunn65} effect (both in hydrogen and helium) towards distant
quasars \cite[]{becker2001}.  Several physical processes have been  
proposed to achieve the total reionisation of the universe. Among 
them, photoionisation by early luminous sources (stars, quasars, dwarf
galaxies) has  recently  been the most favoured. Indeed, it is a rather
simple and natural scenario in which the first luminous sources emit
ionising photons that ionise hydrogen atoms in their vicinities. In
this model, reionisation starts as an inhomogeneous process, and the
total reionisation is reached when the ionised bubbles overlap (due 
to their increasing number and to their expansion). The most recent
observations of very distant quasars ($z\sim 6$) are now suggesting
that we have reached the point where we are starting to observe the period of
inhomogeneous reionisation (IHR) \cite[]{becker2001,gnedin2001}.  As
pointed out by \cite{aghanim96}, inhomogeneous reionisation
can induce secondary anisotropies of the KSZ type, when the ionised
bubbles move with respect to the CMB rest frame. The inhomogeneous,
or patchy, reionisation has been during the few last years the subject 
of numerous recent studies. Among these, we cite for example 
\cite{benson2001},\cite{ciardi2001},\cite{gnedin2001a} and 
\cite{bruscoli2002}\par
We estimate the possible contribution (in terms
of the power spectrum) of the anisotropies due to the IHR by using
the simple model proposed by
\cite{gruzinov98}, which we have generalised to a non-critical
universe (open or flat with non-zero cosmological constant). This is an
empirical model of the IHR based on three parameters: The redshift at
which ionisation starts $z_s$, its duration $\delta z$ and the typical
size of the ionised bubble $R$. In our case, we  add the
constraint that reionisation is completed at $z_i=6$, i.e.  for a given
$z_s$ this fixes $\delta z$. In addition, we set $z_s=10$ which is a
reasonable assumption for the formation of the first objects, i.e. when
ionisation starts. However, this simple model is limited by the fact
that it gives a predicted power spectrum for a specified bubble size
that for illustration we set to 10 Mpc. It therefore introduces an
artificial cut-off at small angular scales. A smaller (larger) size, shifts 
the power spectrum to smaller (larger) angular scales and decreases 
(increases) the amplitude.
\subsection{Ostriker-Vishniac effect}
Along with the IHR effect produced during the first stages of
reionisation and independently of any physical model to explain the
sources of photoionisation, we have to take into account the
well-studied Ostriker-Vishniac effect OV
\cite[]{ostriker86,vishniac87,hu94,dodelson95,jaffe98}
which arises during the linear regime of the cosmological
density fluctuation evolution. Contrary to the IHR, this second order
effect assumes a homogeneous ionisation fraction and relies entirely
on the modulation of the Doppler effect by spatial variations of the
density field which affect the probability of scattering.  Due to its
density squared weighting, it peaks at small angular scales, typically
arcminute scales in low matter-density flat models, and produces $\mu K$ rms
temperature anisotropies.

To estimate its power spectrum, and therefore its contribution to the
temperature anisotropies, we follow the
\cite{jaffe98} approach. The OV effect can be expressed as the
time integral of the velocity field projection along the line of sight
modulated by the density field and weighted by the so-called
visibility function. The visibility function gives the probability of
rescattering of the photons as a function of time (or $z$) and encodes
all the information relevant for  reionisation. The usual approach
is to consider that the visibility function follows a 
Gaussian centred near $z_i$ with a width
corresponding to the interval of reionisation $\delta z$.  This
describes an idealised model for the evolution of the reionisation
process.  However, cosmological simulations
\cite[]{gnedin2000,gnedin2001}, supported by recent observational
data from SLOAN \cite[]{becker2001}, indicate a slow first stage of
reionisation, called the pre-overlap phase, followed by a fast overlap
and post-overlap periods around $z_i \simeq 6$. We thus propose a more
realistic model for the visibility function which shape consists of a
curve with a smooth growing slope at redshifts higher than $z_i$ and 
a steep decrease just before $z_i$.  This approach increases
the amplitude of the OV power spectrum by $50\%$ at all scales and
produces a slight shift of its peak to smaller multipoles (larger
angular scales) as compared to previous works.  This simple but
efficient analytical modeling of the OV effect will depend on the
reionisation redshift, its duration and the cosmological model.
\section{Results}

\begin{figure}[htb]
\begin{center}
\epsfxsize=12cm
\epsfysize=10cm
\epsffile{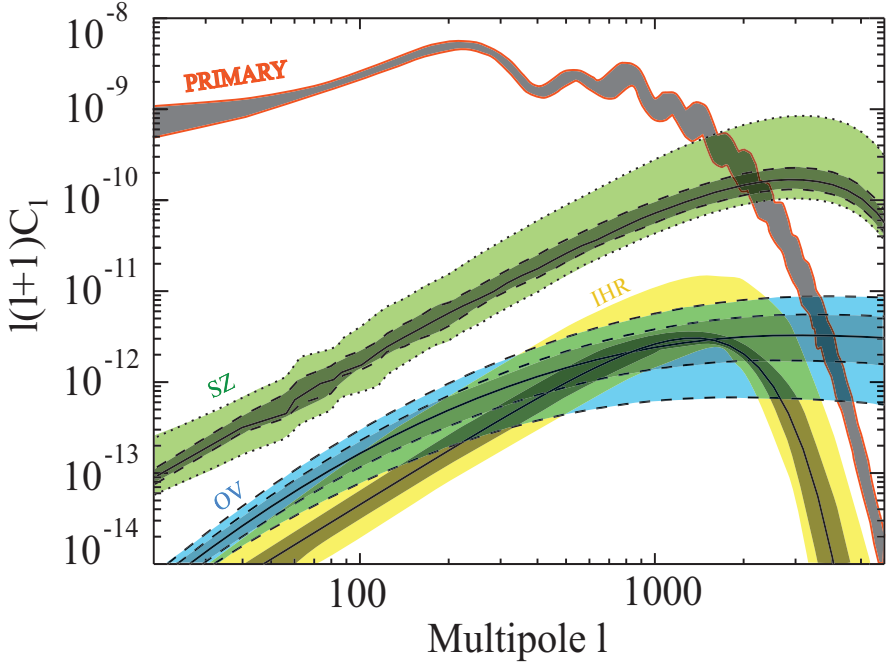}
\end{center}
\caption{Primary and Secondary CMB anisotropies.
The $95 \%$ confidence levels for the primary anisotropies (see text) are
plotted together with the $2\sigma$ envelopes
(light shaded regions) and $1\sigma$ envelopes (heavy shaded regions) for 
the secondary anistropies (see text). The solid line within the
heavy shaded $1\sigma$ envelope represents, for each contribution, 
the power spectrum computed for the ``concordance model''.}
\label{fig:contrib}
\end{figure}
\medskip

The main results of our work are plotted in figure \ref{fig:contrib}.
Constraints on the primary anisotropies are plotted assuming
the template of models described in Section 2.
As we can see, even if multiple oscillations in the CMB data have
not been detected at a level better than $2\sigma$ 
(see, e.g. \cite{douspis2001}) the actual data can constrain the 
region of compatible theoretical models. In particular,
the level of primary anisotropies for $\ell >1000$ can be well
defined, suggesting that  measurements of the power around the
first $3$ peaks constrain the intrinsic parameters of the model template.
Together with the primary anisotropies, we have computed predicted
power spectra for the secondary anisotropies.
For each contributing signal, we give the power spectra for the 
so-called ``concordance model'' ($\Omega_m=0.3$, $\Omega_\Lambda=0.7$, 
$h=0.71$) (solid line within the heavy shaded region). Taking the values 
allowed by the CMB analysis \cite[]{melchiorri2002} and the HST 
measurements of the Hubble parameter \cite[]{freedman2001}, we compute
the power spectra of the $2\sigma$ envelope (with parameters 
$\omega_{cdm}=0.06-0.18$, $h=0.55-0.87$ and $\omega_b=0.018-0.022$),
and of the $1\sigma$ envelope (with parameters 
$\omega_{cdm}=0.09-0.15$, $h=0.63-0.79$ and $\omega_b=0.019-0.021$).
In figure \ref{fig:contrib}, we display both the $2\sigma$ envelopes 
(light shaded regions) and $1\sigma$ envelopes (heavy shaded regions) for 
the different contributions.  

As far as  the SZ effect
is concerned, we emphasize that the majority of present experiments 
cannot disentangle it from the primary CMB due to their 
observing frequencies. 
We therefore calculate the TSZ contribution at two frequencies 
representative of the present experiments: 30 GHz for the
radiometric (DASI and CBI-like) experiments and 150 GHz for the
bolometric (BOOMERanG, MAXIMA and ARCHEOPS-like) experiments. This component
is then added to the KSZ contribution; the total SZ power is then plotted.
The SZ power spectrum peaks between $\ell\sim3200$ and $3600$ depending on 
the cosmological model (but independently of the frequency). Its maximum 
power, in turn, depends more strongly on the set of cosmological parameters
and obviously on the frequency. 
At 30 GHz, the $\ell(\ell+1)C_\ell$ for the ``concordance model'' 
is  $\sim 2.\,10^{-10}$. At 150 GHz, the maximum is about $4.\,10^{-11}$.
We emphasize again that these predictions ought not to be 
taken at face value, but rather as indicative amplitudes for the SZ effect.
The two other secondary effects induce temperature fluctuations which can
be directly expressed in terms of their power spectra regardless of the
frequency. The power spectrum of secondary anisotropies generated 
during the inhomogeneous reionisation peaks in between $\ell\sim 1100$ and 
1800 with an amplitude $\sim 3.\,10^{-12}$ in the ``concordance model''.
As for the OV contribution, the power spectrum peaks around the
multipoles $\ell \sim 2000$ and $4000$, where the primary anisotropies become
subdominant, and its amplitude in the ``concordance 
model'' peaks around $\ell \sim 3000$ with amplitude $3.10^{-12}$.

\begin{figure}[htb]
\begin{center}
\epsfxsize=12cm
\epsfysize=12cm
\epsffile{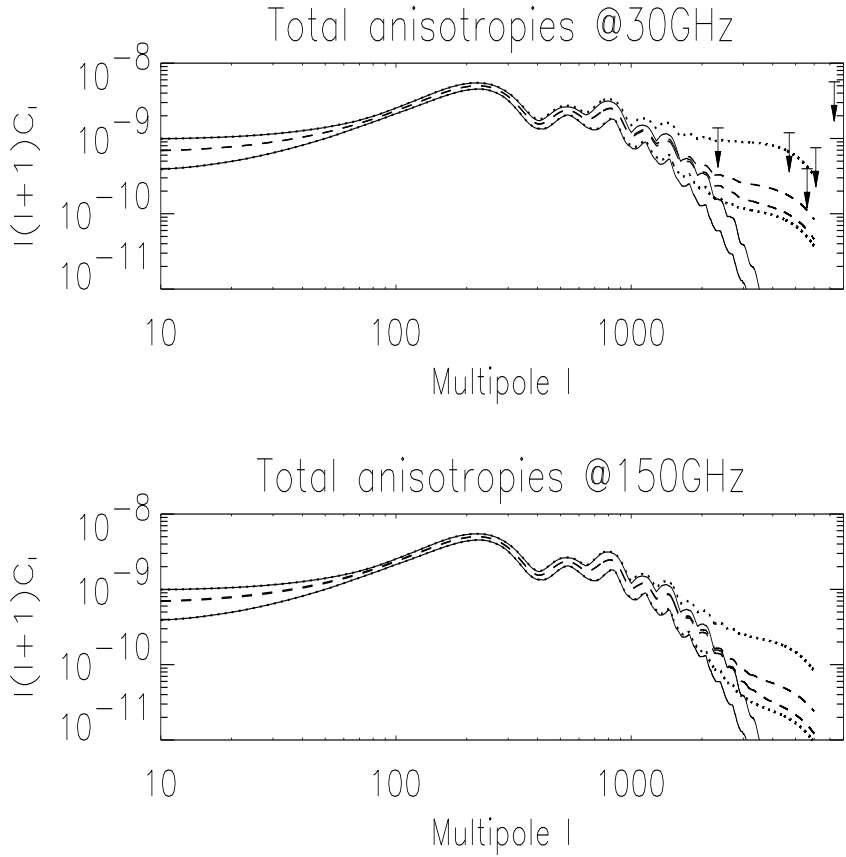}
\end{center}
\caption{Total CMB anisotropies.
The sum of the primary and secondary anisotropies is plotted: Upper panel,
at 30 GHz (for a DASI-like experiment) and lower panel, at 150 GHz (for a 
BOOMERanG-like experiment). 
The solid lines represent the predicted primary CMB region. The dotted and
the dashed lines are respectively the 1 and $2\sigma$ envolpes of the sum
(primary + secondary anisotropies). The upper limits (at 95\%) are 
taken from 
\cite{springel2001} and summarise the results from \cite{subrahmanyan93}, 
\cite{church97}, \cite{ganga97}, \cite{jones97}, \cite{partridge97},
\cite{subrahmanyan98}, \cite{holzapfel2000}.}
\label{fig:cltot}
\end{figure}

It is worth noting that the SZ anisotropies constitute the dominant 
secondary contribution when the TSZ effect is not removed, as  is
the case in the present study. When we compare the power of the 
primary and secondary anisotropies (Fig. \ref{fig:contrib}), we find that
within the $2\sigma$ envelope, the 
secondary contribution equals the
primary power spectrum between $\ell\sim 1600$ and $\ell\sim 2000$. At
30 GHz, the total power spectrum (Fig. \ref{fig:cltot}, upper panel) 
is already significantly modified beyond 
the third acoustic peak and the power spectrum is expected to remain
almost flat down to $\ell=5000$. At 150 GHz, the amplitude of the
secondary contributions (dominated by the TSZ effect in our case) is
smaller. The observed (total) power spectrum in this case (Fig. 
\ref{fig:cltot}, lower panel) is noticeably affected only at the position 
of the sixth acoustic peak. In this case, we do not observe a plateau;
the power slowly decreases with increasing $\ell$s.

\section{Discussion and Conclusions}

In this study, we have examined the sub-arcminute structure of 
the angular power spectrum of the CMB anisotropies.
Using a Bayesian approach, we have shown that under the assumption 
of a wide class of inflationary adiabatic CDM models, the present 
CMB data on degree and sub-degree scales can be used to forecast the
level of anisotropies. We have therefore put strong bounds on the level of
contribution from primary CMB anisotropies to the overall spectrum 
for $\ell >1000$ (on sub-arcminutes scales). These predictions rely
on the experimental results and assume that the present CMB data are
free from any unknown foreground contamination {\it including 
secondary anisotropies}.

We have checked that this hypothesis, and the underlying results, remain
valid when we generalise the foreground to include the secondary 
anisotropies.
We have therefore computed theoretical predictions for three dominant
``candidates'' for the secondary anisotropies: SZ effect, OV effect
and IHR. Future multi-frequency experiments 
observing in the millimetre and sub-millimetre will be able to remove
most of the TSZ contribution. In this case, the remaining SZ signal will be
associated with the KSZ effect. It is one order of magnitude lower than
the values quoted in figure \ref{fig:contrib}, and has the same spectral 
signature as the primary anisotropies. The remaining SZ contribution will 
therefore be larger than, or of the same order of, the OV and IHR 
contributions (which are at the same level). Since all three signals are 
spectrally confused, the multi-frequency experiments will measure 
the sum of the three contributions. The resulting signal is rather
small; it is thus not expected to significantly alter the primary 
anisotropies.  However nowadays,
the present status of the CMB experiments (specifically, their 
frequency coverage)
does not allow us to remove a possible contribution from the TSZ effect of
a cluster population. In this context, the secondary contribution to
the CMB power spectrum is dominated by TSZ. \par
We have shown that such a contribution can be large enough to
significantly modify the CMB power spectrum beyond the third acoustic
peak. This alteration is more pronounced at 30 GHz than at 150 GHz, which
could be the case if the errors bars for the  30 GHz-experiment
are found to be larger than for the 150 GHz-experiments.
The predicted $2\sigma$ envelope for the total power spectrum
(Fig. \ref{fig:cltot}, upper panel, dotted lines) is displayed 
together with the
upper limits (95\%) obtained from small scale experiments quoted in
\cite{springel2001}
\cite[]{subrahmanyan93,church97,ganga97,jones97,partridge97,subrahmanyan98,holzapfel2000}. We note that the upper limits are close to our 
$2\sigma$ region defined by the cosmological parameters allowed by the 
actual constrains. The good consistency between BOOMERanG and DASI data 
(at 150 and 30 GHz respectively) together with the upper limits at small
scales indicates that
{\it the present CMB data can already constrain the contribution from
the secondary effects, TSZ effect in particular}. However, this 
contribution is highly dependent on the cluster abundance and their 
internal physics which can modify our predictions.
In addition, this result has been obtained under a theoretical prior.
The possible presence of ``non-standard'' mechanisms like
primordial voids \cite[]{sakai99,cooray2002,griffiths2002}, 
bumps \cite[]{griffiths2001}, 
scale dependence of the primordial spectral index \cite[]{kosowsky95}
or topological defects \cite[]{avelino2000}, for example, 
can significantly change our conclusions. Fortunately, these mechanisms can 
be distinguished in principle from the models reported here by combination 
with different datasets such as  those from deep redshift galaxy surveys, for
example.
 
Future small-scale CMB data such as  that expected from 
CBI \cite[]{mason2002}, ARCHEOPS \cite[]{amblard2001,benoit2001}, MAP 
and, ultimately, Planck will be helpful for refining our results. The
measurements of the CMB anisotropies after the third peak will 
not only constrain the cosmological model through parameter estimation, but
 will also unable us to probe, via the secondary anisotropies (e.g. SZ), the
formation and evolution of structures\footnote{When we were about to submit
this article,
a similar study came out (Cooray 2002) in which the author
examined  the contributions from secondaries
including IWS and weak lensing for the fiducial $\Lambda$CDM model.}. 

\begin{acknowledgements}
This work was partially supported by the European TMR CMBnet.
PGC is supported by the Funda\c{c}\~{a}o para a Ci\^{e}cia e a Tecnologia.
NA thanks T.F. and the Denys Wilkinson Building for 
``hospitality''.
The authors thank M. Douspis and P. Ferreira for helpful discussion,
and M. Kunz for his comments. We also thank an anonymous referee 
for the comments and suggestions that helped in improving the paper.

\end{acknowledgements}

\end{document}